\magnification 1500
\centerline{\bf Nucleosynthesis  in a simmering universe }
\vskip 3cm
\centerline {Daksh Lohiya, Shobhit Mahajan, Amitabha Mukherjee}
\centerline {Annu Batra}
\centerline {Department of Physics \& Astrophysics, University of Delhi,}
\centerline {Delhi 110 007, India}
\centerline {email: dlohiya@ducos.ernet.in}
\vskip 3cm
\centerline {\bf Abstract}
\vskip 1cm
     Primordial nucleosynthesis is considered a success story of the 
standard big bang (SBB) cosmology. The cosmological and elementary particle
physics parameters are believed to be severely constrained by the 
requirement of correct abundances of light elements. We demonstrate
nucleosynthesis  in a class of models very different from SBB. In these
models the cosmological
scale factor increases linearly with time from the period during 
which nucleosynthesis occurs. It turns out that weak interactions remain in
thermal equilibrium upto temperatures which are two orders of 
magnitude lower than the corresponding temperatures in SBB. Inverse
beta decay of the proton can ensure adequate production of several light 
elements while producing primordial metallicity much higher than that 
produced in SBB. Other attractive features of these models are the 
absence of the horizon, flatness and  initial singularity
problems, consistency with the age of globular clusters and consistent 
relationships between redshift and luminosity distance, angular diameter 
distance and the galaxy number count.

\vfil\eject

	Early universe nucleosynthesis is regarded as a major
``success story'' of the standard big bang (SBB) model. The 
observed light element abundances are believed to severely
constrain cosmological and particle physics parameters. Of late [Steigman, 
1996]  observations have suggested the need for a careful scrutiny and a 
possible revision of the status of SBB nucleosynthesis. While attempts to
reconcile the cosmological abundance of deuterium and the number of 
neutrino generations within the framework of SBB are still on, we feel that
alternative scenarios should be explored. Surprisingly,  a class of models 
radically different  from the standard one can produce the correct amount 
of helium as well as the  metallicity observed in low metallicity objects. 
This paper is a status report on our ongoing efforts to study the 
cosmological implications of a class of models in which the cosmological 
scale
factor $R(t)$ varies linearly with time. 

	A crucial assumption in the standard model is the existence of 
thermal equilibrium at temperatures around $10^{12}K$ or $100 MeV$.
At these temperatures, the universe is assumed to consist of leptons, photons 
and a contamination of nucleons in thermal equilibrium. The ratio of weak 
reaction rates of leptons to the rate of expansion of the universe 
(the Hubble parameter) below $10^{11}K$ (age $\approx .01 $secs) 
goes as [see eg., Weinberg, 1972]

$$ {\sigma n_l\over H} \approx ({T\over 10^{10}K})^3 \eqno{(1)}$$

At these temperatures, the small nucleonic  
contamination begins  to shift towards more protons and fewer
neutrons because of the n-p mass difference. 
By $10^{10}K$ i.e. $T_9 = 10$, 
the neutrinos  decouple. The distribution function 
of the $\nu$'s however
maintains a Planckian profile as the universe expands. At 
$ 5*10^{9}K$ (age of about  4 secs), $e^+,e^-$ pairs
annihilate. The neutrinos having decoupled, all the entropy of
the $e^+, e^-$ before annihilation, goes to heat up the photons -
giving the photons some 40\% higher temperature than the temperature
corresponding to the neutrino Planckian profile. The decoupling
of the neutrinos and the annihilation of the $e^+, e^-$ ensures the
rapid fall of the neutron production rate 
$\lambda(p\longrightarrow n)$ in comparison to the 
expansion rate of the universe. The n/p ratio freezes to about 
1/5 at this epoch. 
This ratio now falls slowly on account of the decay of
free neutrons. Meanwhile nuclear reactions and photo -
disintegration of light nuclei ensure a dynamic buffer of light elements
with abundances roughly determined by nuclear statistical 
equilibrium (NSE). Depending on the baryon-entropy ratio, at a critical 
temperature around $T_9 = 1$, the deuterium concentration is large enough 
for efficient evolution of a whole network of reactions leading up to the 
formation of the most stable light nucleus, viz. $He_4$. At slightly lower 
temperatures, the deuterium depletion rate becomes small compared 
to the expansion rate [see eg., Kolb \& Turner, 1989] resulting in residual 
abundances of deuterium and $He_3$.  Elaborate numerical codes have 
been developed by Wagoner and Kawano [WK] [1988] to describe the evolution
of this phase. While the predicted abundances of deuterium, helium - 3,
helium - 4 and lithium - 7 are believed to be consistent with observations,
one does not see any astrophysical object with metallicity (abundance
of lithium - 8 and heavier elements) as low as that predicted by
primordial synthesis alone. The oldest objects are believed to be
globular clusters. The metallicity reported in these systems is much
higher than accounted for by SBB and much too low in comparison with
that found in the atmosphere of population I stars and interstellar
gas. Consistency of the light element abundances 
in SBB, moreover, is ensured only if the baryonic matter density is some 
two orders of magnitude less than the closure density. This is regarded 
as a respite in SBB. Using the rest of the (non-baryonic) matter in a 
suitable combination of hot and cold dark matter (with possibly a 
small cosmological constant also thrown in) to build up 
large scale structures in cosmology has developed into an industry. 
The current status is far  from satisfactory [See eg., Ostriker et al 1995].
In particular, the age estimates of globular clusters are uncomfortably high
in comparison with the age of the universe as set by 
conservative estimates.

	Motivated by the above, we explore the possibility of obtaining a 
consistent scenario for nucleosynthesis in a class of models which are
radically different from the standard one. In particular, we consider a 
cosmological model in which, at the epoch when 
$T \approx 10^{12}K$ and thereafter, the scale factor $R(t)$ increases 
as $t$. The linear evolution of the scale factor ensures a horizon-free 
cosmology. 
We shall describe later how this can be ensured. We shall refer 
to $t$ as the age of the universe. The present value of the scale parameter
and the present epoch  $t_o$ are exactly determined by the present
Hubble constant $H_o = 1/t_o$. The scale factor and the temperature of
radiation are related by  $RT =$ constant. In such a model, the age of the 
universe when $T \approx 10^{10}K$ would be of the order of a few years. The
universe takes some $10^2$ to $10^3$ years to cool from $10^{10}K$ 
to $10^8K$. The rate of expansion of the universe is about $10^7$ times 
slower than the corresponding rates for the same temperature in standard 
cosmology. This makes a crucial difference and in fact ensures that the  
standard story does not go through. 

      The process of the neutrinos falling out of thermal equilibrium, for 
example, is determined by the rate of $\nu$ production per charged lepton:
$$ \sigma_{wk} n_l/c^6 \approx g_{wk}\hbar^{-7}(kT)^5/c^6\eqno{(2)}$$
and the expansion rate of the universe [H = 1/t]. Here
$g_{wk} \approx 1.4*10^{-45}$ erg- $cm^3$. For 
$kT > m_\mu, T > 10^{12}K$
$$ \sigma_{wk} n_l/H \approx [{T\over {1.62*10^{8}K}}]^4 
\eqno{(3)}$$
Here we have normalized the value of $RT = tT =$ constant
from the value $H_o$ = 55 km/sec/Mpc for the Hubble constant. 
corresponding to $t_o \approx 12*10^9$ years. Increasing 
$H_o$ by a factor of 2 would merely lead to a change in the 
denominator on the right side of eqn. 3 to  $1.8*10^{8}K$.
When $kT < m_\mu$, the number density of muons is reduced by 
a factor $[exp(-m_\mu/kT)]$. Consequently, the rates of 
weak interactions involving muons get suppressed to
$$ \sigma_{wk} n_l/H \approx [{T\over {1.62*10^{8}K}}]^4 
exp[-{10^{12}K\over T}] \eqno{(4)}$$

The corresponding rates in the standard model are:
$$ \sigma_{wk} n_l/H \approx [{T\over {10^{10}K}}]^3 \eqno{(5)}$$
for $kT > m_\mu$, and
$$ \sigma_{wk} n_l/H \approx [{T\over {10^{10}K}}]^3 
exp[-{10^{12}K\over T}] \eqno{(6)}$$
for $kT < m_\mu$. This would lead to the weak interactions
maintaining the $\nu$'s in thermal equilibrium to temperatures down 
to $1.62*10^{8}K$. This would then imply that the entropy
released from the $e^+ e^-$ annihilation would heat up 
all the particles in equilibrium. Both the neutrinos and
the photons would get heated up to the same temperature.
The temperature then scales by $RT =$ constant as the 
universe expands. The relic neutrinos and the photons 
(the CMBR) would therefore have the same Planckian profile
($T \approx 2.7K$) at present. This is in marked contrast
to the standard result wherein the neutrino temperature
is predicted to be lower than the photon temperature.
The nuclear reaction rates are simply given by the 
expressions [Weinberg, 1972]:
$$ \lambda(n\longrightarrow p) = A\int(1 - {m_e^2\over 
(Q+q)^2})^{1/2}(Q+q)^2q^2dq$$
$$*(1 + e^{q/kT})^{-1}(1 + e^{-(Q+q)/kT})^{-1}\eqno{(7)}$$
$$ \lambda(p\longrightarrow n) = A\int(1 - {m_e^2\over 
(Q+q)^2})^{1/2}(Q+q)^2q^2dq$$ 
These rates have the ratio:
$${\lambda(p\longrightarrow n)\over 
{\lambda(n\longrightarrow p}} = exp(-{Q\over kT})\eqno{(8)}$$
The rate of expansion of the universe at a given 
temperature being much smaller than that in the standard scenario, 
the nucleons are expected to be in thermal equilibrium with the ratio $X_n$ 
of neutron number to the total number of all nucleons
given by:
$$ X_n = {\lambda(p\longrightarrow n)\over 
{\lambda(p\longrightarrow n) + \lambda(n\longrightarrow p)}}
= [1 + e^{Q/kT}]^{-1}\eqno{(9)}$$
   
    However the conditions of NSE would still hold. A buffer
of light elements would emerge as before. The baryonic content
of the universe at 
$T_9 \approx 1$ is constituted by protons (mainly), some neutrons (less 
than 1\% ) and a buffer of light elements in NSE. The strength of the
buffer is enhanced by fresh neutron formation by the inverse beta
decay of the proton and its capture into the buffer by the pn
reaction. The buffer depletes by either: (i) the photodisintegration 
of any light element constituting the buffer followed by the decay 
of the resulting neutron before it can be recaptured into the 
buffer by the pn reaction; or (ii) the formation of $He_4$ which
is the most stable nucleus at these temperatures. Once helium 
formation becomes more efficient than neutron decay, all subsequent 
neutrons formed would precipitete into $He_4$.
This critical epoch is sensitive to the baryon-entropy
ratio. If the ratio of number of 
protons 
that convert into neutrons after that epoch to the total baryon number of
the universe is roughly 1/8, we would get the observed $\approx 25\%$ He. 
This simply translates 
into an appropriate requirement on the baryon-entropy ratio. 

       We have modified the (WK) numerical code outlined by Kawano
to suit the taxing requirements of the much stiffer rate equations
that we encounter in our slowly evolving universe. To get 
convergence of the rate equations  for 26 nuclides and a network of
88 reactions, as given in Kawano, we executed some 500 iterations 
at each time step. We have incorporated an additional (89th) reaction:
the pp reaction 
$$ p + p \longrightarrow D + e^+ + \nu \eqno{(10)}$$ 
As a consequence of this reaction, the lifetime of protons is around  
$10^{10}$ years in the core of a typical star at temperatures of the 
order of $T_9 \approx .01$ and 
densities some $100 gm cm^{-3}$. The contribution of this reaction
in the few minutes that the universe has temperature $\approx T_9 = 1$
and density $< 1 gm cm^{-3}$ is negligible in SBB
but in our model where the expansion rate is some $100 yrs^{-1}$,
it gives a substantial contribution.  
The results for different values of $\eta$ are described in
table I.  We find consistency with the $He_4$ abundances for
$\eta \approx 10^{-8} $. The high metallicity produced is 
also a consequence of the slow expansion in this model.

    To get the observed abundances of the light elements, one would
have to rely on nucleosynthesis by secondary explosions of
supermassive objects [Wagoner 1969]. 
We feel we may be able to dynamically account
for such explosions within the framework of models we shall outline later.

	The expansion rate in this model does not depend on the background 
density and thus the abundances are independent of the number of neutrino 
species. The age of this universe (defined as the time elapsed from the hot
epoch to the present) would be exactly $50\%$ higher than the
SBB age determination from the inverse of the present Hubble
parameter. If one goes by the value of 80 km/sec/Mpc currently quoted for 
the Hubble parameter, our model would accommodate 15 giga-year
old objects much more comfortably than SBB where one merely advocates
disbelief in single-$\sigma$ error bars [Longair 1995]. 

	 We now address the issue of realising the linear evolution 
within the framework of a Friedman cosmology. Such an evolution
can be accounted for in a universe dominated by `K - matter'$^7$ for which 
the density scales as $R^{-2}$. 
The Hubble diagram (luminosity distance-redshift 
relation), the angular diameter distance - redshift relation and the galaxy 
number count-redshift relations do not rule out such  
a ``coasting'' cosmology [Kolb, 1989; Sethi et al 1996].
However, if one requires
this matter to dominate even during the nucleosynthesis era, the K -
matter would almost close the universe. There would be hardly any
baryons. An alternative way of achieving a linear  
evolution of the scale factor is an effective
Einstein theory with a repulsive  gravitational constant at long distances.
Such possibilities follow from effective gravitational actions
that have been considered in the past. Ellis and Xu [1995], for 
example, considered a fourth order theory with action:
$$ S = \int d^4x\sqrt{-g}[\alpha R^2 - \beta R] \eqno{(11)}$$
In the weak field approximation, the effective Newtonian potential
in such a theory is:
$$\phi = - {a\over r} + b{exp(-\mu r)\over r} \eqno{(12)}$$
For $\mu r << 1$ we can have a canonical effective attractive theory. At 
large distances, the effective potential is dominated by the first term
alone - corresponding to repulsive gravity. A similar possibility occurs
in the conformally invariant proposal of Manheim and Kazanas[1990].
Choosing the gravitational action to be the square of the
Weyl tensor gives rise to an effective induced action:

$$ S = \int d^4x\sqrt{-g}[\alpha C^2 - \beta R]\eqno{(13)}$$

The dynamics of a conformally flat FRW 
metric is driven by the anomalous repulsive term alone. Canonical
attractive domains occur in the model as non - conformally flat 
perturbations in the FRW spacetime. Repulsive long range gravitation
with spherical attractive domains also occur in a variant of a model by 
Zee [1982]:

$$ S = \int d^4x\sqrt{-g}[-\epsilon\phi^2R + 
{1\over 2}(\partial_\mu\phi)(\partial^\mu\phi)
+ V(\phi)+ \beta_{ind}R + \Lambda_{ind} + L_m]\eqno{(14)}$$

Here $\beta_{ind}$ and $\Lambda_{ind}$ are induced gravitational
and cosmological constants, $L_m$ the action for the 
rest of the matter fields and $V(\phi)$ the effective potential           
for the scalar field.
In these models, static non- topological soliton solutions exist. For such
solutions, $\phi$ is constant inside a sphere and rapidly 
goes to zero near its surface.
These solutions would have an effective canonical attractive
gravitation in their interior and have repulsive gravitation
outside. 

	 We have been exploring the possibility that 
such non-topological domains - gravity balls (g-balls) -  
of the size of a typical galactic halo (or larger) play 
an essential role in cosmology. The formation of large scale structure  
by the splitting of gravity balls, the growth of density perturbations  
and gravitational lensing are some of the areas which are the subject
of our continuing investigation. In this context, we find the results of 
this paper quite encouraging.

\vskip 2cm

Acknowledgment: Helpful discussions with Prof. Jim Peebles in ICGC 1995, Pune,
India, are
gratefully acknowledged.
 
\vfil\eject

\centerline{\bf References}

\item{1.} Batra, A.`Studies on alternative theories' M. Sc.dissertation,
University of Delhi 1995, unpublished. Also presented as a poster in
ICGC 1995, Pune, India.

\item{2.} Ellis, G.F.R. and Xu M., 1995 (private communication).

\item{3.} Kawano L., 1988, FERMILAB - PUB -88/34 and references therein.

\item{4.} See for example Kolb, E.W. and Turner M.S, 1990, The Early Universe,
Addison Wesley.

\item{5.} Kolb E. W.,
1989, Ast. Phys. J. 344, 543.

\item{6.}Lohiya D., [1995], ``Non - topological solitons in
non - minimally coupled scalar fields: theory and consequences''
Ast. and Sp. Sci. [to be published].

\item{7.} Longair M., 1995, talk given in ICGC, Pune

\item{8.} Manheim P. and Kazanas D. [1990], Gen. Rel \& Grav. 22, 289.

\item{9.} Ostriker J.P., Babul A., Weinberg D.H. and Dekel, [1993], Testing 
the 
gravitational instability hypothesis, Univ. Princeton preprint.

\item{10.}Sethi M. \& Lohiya, D. 1996, ``Observational consequenses of an
asymptotically coasting cosmology'' in preparation

\item{11.} Steigman G., 1996, ``Big - Bang Nucleosynthesis'', Astro-ph/9601126

\item{12.} Wagoner R. V., Ap. J. Supp., 162, 18, 247

\item{13.} Weinberg S., Gravitation and Cosmology, John Wiley \&
sons, [1972].

\item{14.} Zee [1982], in ``Unity of forces in nature''
Vol II, ed. A. Zee, P 1082, World Scientefic Pubs.

\item{15.} Zee  A., [1979] Phys. Rev. Lett., 42, 417; Phys. Rev. Lett. 44, 
703,
1980.

\vfill
\eject

\vskip 3cm

\centerline {\bf {TABLE I}}
\vskip 1cm
\centerline{
Abundances of Some Light Elements and Metals.} 

\vskip 2cm

\settabs 7 \columns
\+\bf$\eta$&$\bf^2 H$&$\bf^3 H$&
$\bf^3 He$&$\bf^4 He$&$\bf^7 Be$&$\bf^8 Li$ \& above \cr
\smallskip
\+$(10^{-9})$&$(10^{-18})$
&$(10^{-25})$&$(10^{-14})$&$(10^ {-1})$&$(10^{-11})$&$(10^{-9})$ \cr
\smallskip
\+$9.0$&$2.007$&$1.25$&$8.65$&$2.03$&$1.39$&$8.06$ \cr
\+$9.1$&$2.008$&$1.26$&$8.63$&$2.06$&$1.32$&$8.63$ \cr
\+$9.2$&$2.009$&$1.26$&$8.60$&$2.10$&$1.23$&$9.35$ \cr
\+$9.3$&$2.010$&$1.27$&$8.59$&$2.11$&$1.19$&$9.75$ \cr
\+$9.4$&$2.014$&$1.26$&$8.56$&$2.15$&$1.11$&$10.66$ \cr
\+$9.5$&$2.015$&$1.27$&$8.50$&$2.18$&$1.05$&$11.41$ \cr
\+$9.6$&$2.016$&$1.28$&$8.52$&$2.19$&$1.01$&$11.88$\cr
\+$9.7$&$2.017$&$1.28$&$8.49$&$2.22$&$0.96$&$12.69$ \cr
\+$9.8$&$2.020$&$1.29$&$8.47$&$2.25$&$0.91$&$13.51$ \cr
\+$9.9$&$2.020$&$1.29$&$8.45$&$2.28$&$0.86$&$14.47$ \cr
\+$10.0$&$2.020$&$1.30$&$8.43$&$2.30$&$0.83$&$15.19$ \cr

\bigskip
\noindent
\+Initial Temperature   $10^{11}K$ \cr
\+Final Temperature  $10^7K$ \cr 
\+No. of iterations at each step  550 \cr

\bye